\documentclass[letterpaper,twocolumn,superscriptaddress,floatfix]{revtex4}
\usepackage[latin1]{inputenc}
\usepackage{bm}
\usepackage{multirow,amssymb,amsbsy,amsmath}
\usepackage{graphicx}
\usepackage{verbatim}
\makeatletter
\usepackage{pifont}
\makeatother

\begin{document}

\title{Experimental control of the transition from Markovian to non-Markovian dynamics of open quantum systems}

\author{Bi-Heng Liu}
\affiliation{Key Laboratory of Quantum Information, University of Science and Technology of China, CAS, Hefei, 230026, China}

\author{Li Li}
\affiliation{Key Laboratory of Quantum Information, University of Science and Technology of China, CAS, Hefei, 230026, China}

\author{Yun-Feng Huang }
\affiliation{Key Laboratory of Quantum Information, University of Science and Technology of China, CAS, Hefei, 230026, China}

\author{Chuan-Feng Li}
\email{cfli@ustc.edu.cn}
\affiliation{Key Laboratory of Quantum Information, University of Science and Technology of China, CAS, Hefei, 230026, China}

\author{Guang-Can Guo}
\affiliation{Key Laboratory of Quantum Information, University of Science and Technology of China, CAS, Hefei, 230026, China}

\author{Elsi-Mari Laine}
\affiliation{Turku Centre for Quantum Physics, Department of Physics and Astronomy, University of
Turku, FI-20014 Turun yliopisto, Finland}

\author{Heinz-Peter Breuer}
\affiliation{Physikalisches Institut, Universit\"at Freiburg,
Hermann-Herder-Strasse 3, D-79104 Freiburg, Germany}

\author{Jyrki Piilo}
\email{jyrki.piilo@utu.fi}
\affiliation{Turku Centre for Quantum Physics, Department of Physics and Astronomy, University of
Turku, FI-20014 Turun yliopisto, Finland}

\date{May 5, 2011}

\maketitle

{\textbf{Realistic quantum mechanical systems are always exposed to an external environment. The presence of the environment often gives rise to a Markovian process in which the system loses information to its surroundings. However, many quantum systems exhibit a pronounced non-Markovian behavior in which there is a flow of information from the environment back to the system, signifying the presence of quantum memory effects \cite{Breuer2007, Fleming, Burghardt, Rebentrost, Paternostro}. The environment is usually composed of a large number of degrees of freedom which are difficult to control, but some sophisticated schemes for modifying the environment have been developed \cite{Wineland}. The physical realization and control of dynamical processes in open quantum systems plays a decisive role, for example, in recent proposals for the generation of entangled states \cite{Diehl,Krauter,Cho}, for schemes of dissipative quantum computation \cite{Verstraete}, for the design of quantum memories \cite{Pastawski} and for the enhancement of the efficiency in quantum metrology \cite{Goldstein}. Here we report an experiment which allows through selective preparation of the initial environmental states to drive the open system from the Markovian to the non-Markovian regime, to control the information flow between the system and the environment, and to determine the degree of non-Markovianity by direct measurements on the open system.}}

The standard approach to the dynamics of open quantum systems employs the concept of a quantum Markov process which is given by a semigroup of completely positive dynamical maps and a corresponding quantum master equation with a generator in Lindblad form \cite{Gorini,Lindblad}. Very recently, a toolbox for the engineering of such quantum Markov processes in a multi-qubit system of trapped ions has been realized experimentally \cite{Barreiro} and technological developments have also allowed experimental studies of quantum correlations in open systems \cite{Li1,Li2}.
Within a microscopic approach quantum Markovian master equations are usually obtained by means of the Born-Markov approximation which presupposes a weak system-environment coupling and several further, mostly rather drastic approximations. However, in many processes occurring in nature these approximations are not applicable, a situation which occurs, in particular, in the cases of strong system-environment couplings, structured and finite reservoirs, low temperatures, as well as in the presence of large initial system-environment correlations. In the case of substantial quantitative and qualitative deviations from the dynamics of a quantum Markov process one often speaks of a non-Markovian process, implying that the dynamics is governed by significant memory effects. Quite recently important steps towards the development of  a general consistent theory of non-Markovian quantum dynamics have been made which try to rigorously define the border between Markovian and non-Markovian quantum evolution and to quantify memory effects in the open system dynamics \cite{Wolf,BLP,BLP2,RHP}.

The measure for quantum non-Markovianity constructed in \cite{BLP} is based on the idea that memory effects in the open system dynamics can be characterized in terms of the flow of information between the open system and its environment. It has been used recently, e.g., to describe this information flow in the energy transfer dynamics of photosynthetic complexes \cite{Rebentrost,Fleming}, and to characterize memory effects of the dynamics of qubits in spin baths \cite{Paternostro}. Here, we present the results of an experiment which enables through a careful preparation of the initial system-environment states and quantum state tomography not only to control and to monitor the transition from Markovian to non-Markovian quantum dynamics, but also the direct determination of this measure for quantum non-Markovianity.

Quantum memory effects are quantified by employing the trace distance $D(\rho_1,\rho_2)=\frac{1}{2}{\mathrm{tr}}|\rho_1-\rho_2|$
between two quantum states $\rho_1$ and $\rho_2$. This quantity can be interpreted as a measure for the distinguishability of the two states \cite{Hellstrom,Holevo,Hayashi}. Markovian processes tend to continuously reduce the distinguishability of physical states, which means that there is a flow of information from the open system to its environment. In view of this interpretation the characteristic feature of a non-Markovian quantum process is the increase of the distinguishability, i.e. a reversed flow of information from the environment back to the open system. Through this recycling of information the earlier states of the open system influence its later states, which expresses the emergence of memory effect in the open system's dynamics \cite{BLP,BLP2}.

On the basis of this physical picture one can construct a general measure for the degree of non-Markovianity of a quantum process given by some dynamical map $\Phi_t$ which maps the initial states $\rho(0)$ of the open system to the corresponding states $\rho(t)=\Phi_t\big(\rho(0)\big)$ at time $t$. The full time evolution of the open system over some time interval from the initial time $0$ to the final time $T$ is then given by a one-parameter family $\Phi=\{\Phi_t \mid 0 \leq t \leq T \}$ of dynamical maps. We define the rate of change of the trace distance by $\sigma(t,\rho_{1,2}(0)) = \frac{d}{dt}D(\rho_1(t),\rho_2(t))$. The measure ${\mathcal{N}}(\Phi)$ for the non-Markovianity of the process is then defined by
\begin{equation} \label{MEASURE}
 {\mathcal{N}}(\Phi) = \max_{\rho_{1,2}(0)} \int_{\sigma > 0}
 dt \; \sigma(t,\rho_{1,2}(0)).
\end{equation}
Here, the time-integration is extended over all subintervals
of $[0,T]$ in which the rate of change of the trace distance $\sigma$ is positive, and the maximum is taken over all pairs of initial states. The quantity (\ref{MEASURE}) thus measures the maximal total increase of the distinguishability over the whole time-evolution, i.e., the maximal total amount of information which flows from the environment back to the open system.

\begin{figure}[tb]
\centering
\includegraphics[width=0.4\textwidth]{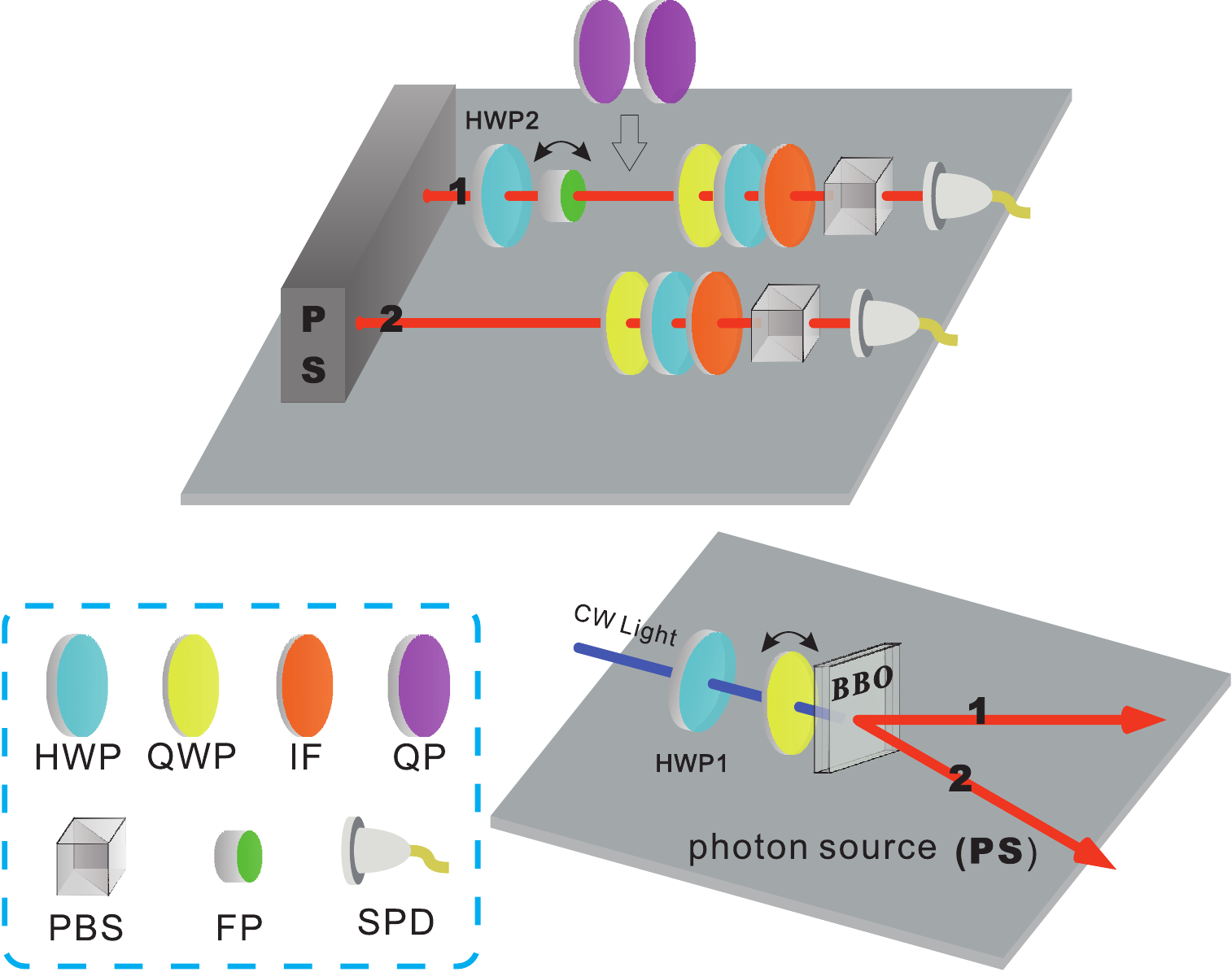}
\caption{\label{Fig:1} The experimental setup. Here, the abbreviations of the components are: HWP -- half wave plate, 
QWP -- quarter wave plate, IF -- interference filter, QP -- quartz plate, PBS --  polarizing beamsplitter, 
FP -- Fabry-Perot cavity, and  SPD -- single photon detector.  }
\end{figure}

In our experiment the open quantum system is provided by the polarization degree of freedom of photons coupled to the frequency degree of freedom representing the environment. The experimental setup is shown in Fig.~\ref{Fig:1}. An ultraviolet Argon-Ion laser is used to pump two  $0.3 \textrm{mm}$ thick BBO crystals cut for type I down conversion process to generate arbitrary pure two-qubit states. A fused silica plate ($0.04 \textrm{mm}$ thick and coated with partial reflecting coating on each side, with about $85\%$ reflection probability at $702 \textrm{nm}$) is used as a FP cavity. The cavity is mounted on a rotator which can be tilted in the horizontal plane.
A $4$nm (full width at half maximum) interference filter is placed after the FP cavity to filter out at 
most two transmission peaks. The corresponding interference filter in the other arm is $10$nm.
The polarization and frequency degrees of freedom are coupled in a quartz plate in which different evolution times are realized by varying the thickness of the plate. 
A polarizing beam splitter together with a half-wave plate and a quarter-wave plate is used as a photon state analyzer.

\begin{figure}[tb]
\centering
\includegraphics[width=0.4\textwidth]{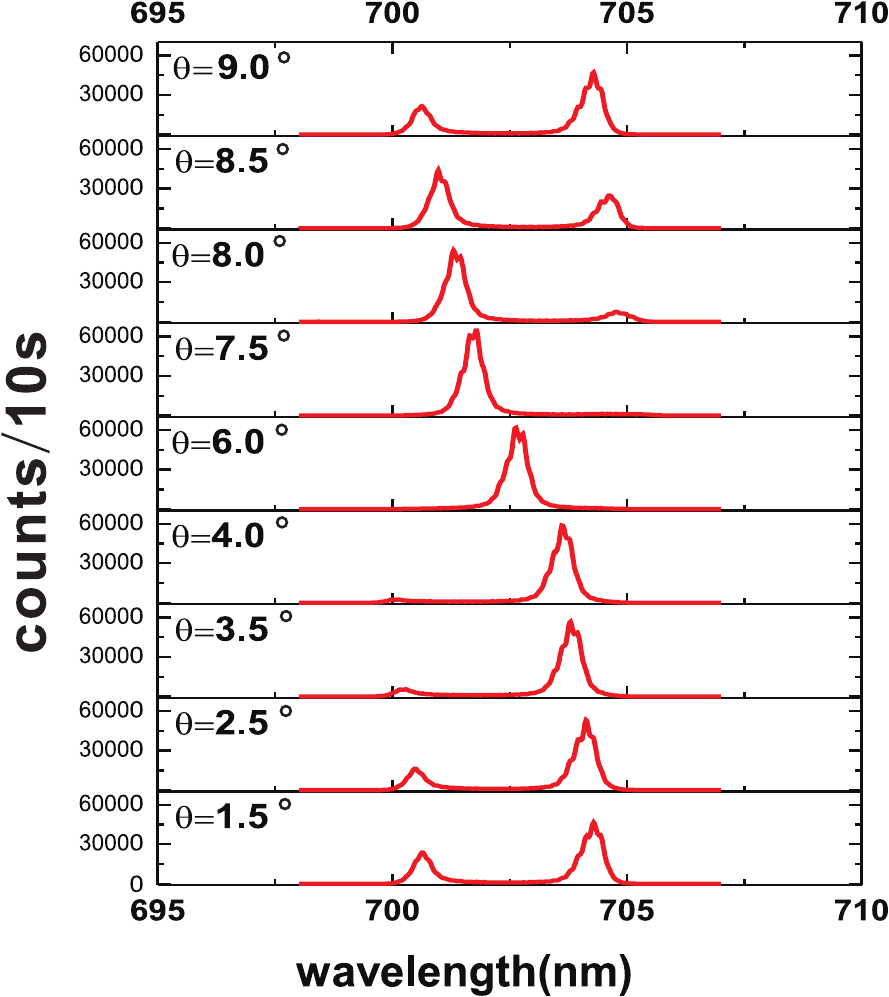}
\caption{\label{Fig:2} The frequency spectrum of the initial state for various values of the tilting angle $\theta$.}
\end{figure}

The half wave plate HWP2 and the tilted FP cavity are used to prepare the initial one-photon states
$|\psi_{1,2}(0)\rangle = |\varphi_{1,2}\rangle \otimes |\chi\rangle$, where
\begin{equation} \label{init-pair}
 |\varphi_{1,2}\rangle = \frac{1}{\sqrt{2}}
 \left( |H\rangle \pm |V\rangle \right),
\end{equation}
with $|H\rangle$ and $|V\rangle$ denoting the horizontal and
the vertical polarization state, respectively. The environmental state $|\chi\rangle = \int d\omega f(\omega) |\omega\rangle$ involves the amplitude $f(\omega)$ for the photon to be in a mode with frequency $\omega$, which is normalized as $\int d\omega |f(\omega)|^2 = 1$. The form of the frequency distribution $|f(\omega)|^2$ and thus the initial state of the environment can be controlled by the tilting angle $\theta$ of the FP cavity. Figure \ref{Fig:2} shows how $\theta$ determines the structure of the frequency spectrum and, thus, the environmental initial state $|\chi\rangle$. By changing the initial state of the environment in the experiment we modify the open system dynamics in a way which allows us to observe transitions between Markovian and non-Markovian quantum dynamics.

\begin{figure}[tb]
\begin{center}
\includegraphics[width=0.4\textwidth]{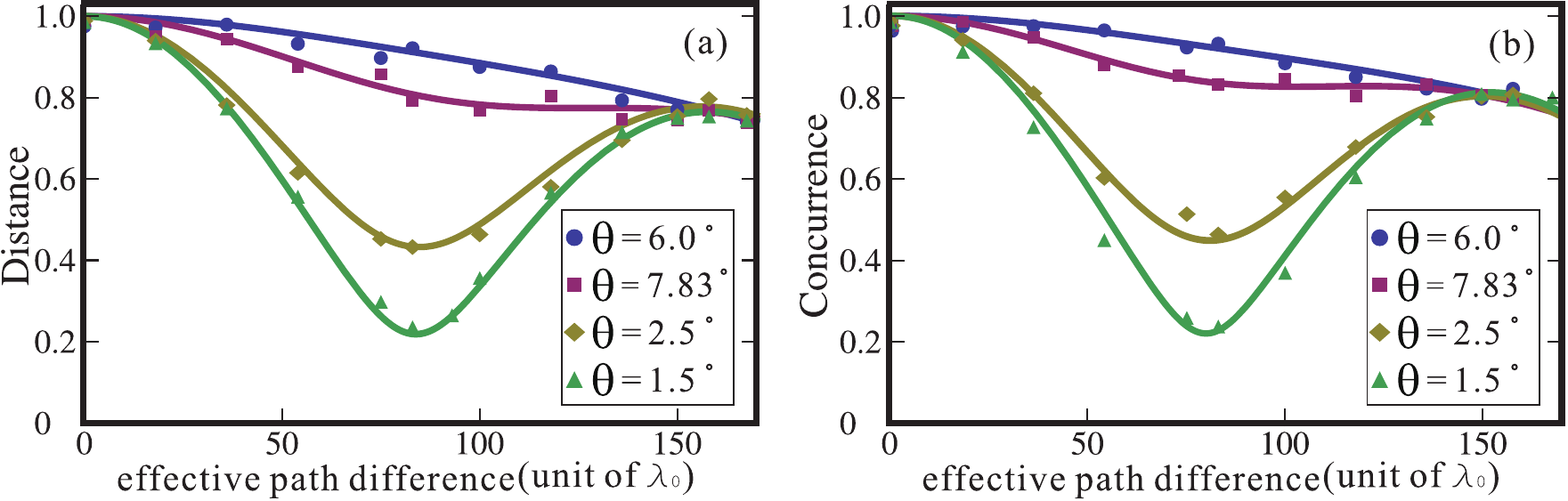}
\caption{\label{Fig:3}
The trace distance (a) and the concurrence between the system and the ancilla (b) as a function of the effective path difference for four different values of the tilting angle $\theta$. The solid lines represent the theoretical predictions for $\sigma=1.8\times10^{12}\textrm{Hz}$ and $\Delta\omega=1.6\times10^{13}\textrm{Hz}$. The effective path difference is equal to $\Delta n L$ and $\lambda_0=702$nm.
The experimental error bars due to the counting statistics are smaller than the symbols.}
\end{center}
\end{figure}

In the first version of the experiment, HWP1 is fixed at zero degree to generate a two-photon state. Photon 2 is directly detected in detector SPD in the end of  arm 2 as a trigger of photon 1. Photon 1 first passes HWP2, preparing it in the state $|\varphi_1\rangle$ or $|\varphi_2\rangle$. The subsequent interaction between polarization and mode degrees of freedom in the quartz plate is described by a quantum dynamical map $\Phi_t$ acting on the open system, where the interaction time $t$ is connected to the variable length $L$ of the quartz plate by means of $t=L/c$. Finally, a full state tomography is carried out in detector SPD in the end of the arm 1 to determine the polarization state $\rho_{1,2}(t) = \Phi_t\left(|\varphi_{1,2}\rangle\langle\varphi_{1,2}|\right)$ of photon 1. This allows the direct experimental determination of the trace distance $D(\rho_1(t),\rho_2(t))$ between the two possible one-photon states after a certain interaction time $t$ controlled by the length $L$ of the quartz plate. Experimental results for four different values of the tilting angle $\theta$ of the FP cavity are shown in Fig.~\ref{Fig:3}(a). We clearly observe a crossover from a monotonic to a non-monotonic behavior of the trace distance as a function of time, i.e., a transition from a Markovian to a non-Markovian dynamics of the polarization degree of freedom of the photon.

The experimental results admit a simple theoretical analysis which is based on the fact that the time evolution in the quartz plate may be described by the unitary operator $U(t)$ which is defined by
\begin{equation*}
 U(t)|\lambda\rangle \otimes |\omega\rangle
 = e^{in_{\lambda}\omega t} |\lambda\rangle \otimes |\omega\rangle,
\end{equation*}
where $n_{\lambda}$ represents the refraction index for light with polarization $\lambda=H,V$. The presence of the quartz plate thus leads to the decoherence of superpositions of polarization states,  which is due to a nonzero difference $\Delta n = n_V-n_H$ in the refraction indices of horizontally and vertically polarized photons. The corresponding dynamical map $\Phi_t$ takes the form:
\begin{equation*}
 \Phi_t: \left\{
 \begin{array}{ll}
 |H\rangle\langle H| & \mapsto |H\rangle\langle H|, \\
 |V\rangle\langle V| & \mapsto |V\rangle\langle V|, \\
 |H\rangle\langle V| & \mapsto \kappa^*(t) |H\rangle\langle V|, \\
 |V\rangle\langle H| & \mapsto \kappa(t) |V\rangle\langle H|,
 \end{array} \right.
\end{equation*}
where the decoherence function $\kappa(t)$ is given by the Fourier transform of the frequency distribution,
\begin{equation*}
 \kappa(t) = \int d\omega |f(\omega)|^2 e^{i\omega\Delta n t}.
\end{equation*}
With the help of these relations it is easy to show that the trace distance corresponding to the initial pair of states (\ref{init-pair}) is equal to the modulus of the decoherence function,
\begin{equation} \label{OPTIMAL-DISTANCE}
 D(\rho_1(t),\rho_2(t)) = |\kappa(t)|.
\end{equation}
In the experiment the transition from Markovian to non-Markovian dynamics is observed through variation of the tilting angle $\theta$ of the FP cavity. As illustrated in Fig.~\ref{Fig:2}, all frequency distributions are very well approximated by a sum of two Gaussians centered at frequencies $\omega_k$ with amplitudes $A_k$ and equal widths $\sigma$, which yields
\begin{equation}
 |\kappa(t)| = \frac{e^{-\frac{1}{2}\sigma^2(\Delta n t)^2}}{1+A}
 \sqrt{1+A^2+2A\cos(\Delta\omega \cdot \Delta n t)},
 \label{kappa}
\end{equation}
where $A_1 = \frac{1}{1+A}$, $A_2 = \frac{A}{1+A}$ and $\Delta\omega=\omega_2-\omega_1$. The tilting angle of the cavity is relatively small and thus the distance $\Delta\omega$ between the peaks remains approximately constant. Therefore, the only relevant parameter controlling the transition is the relative amplitude $A$. Equation \eqref{kappa} yields an excellent approximation of the experimental data (see the continuous lines in Fig.~\ref{Fig:3}).

\begin{figure}[tb]
\centering
\includegraphics[width=0.4\textwidth]{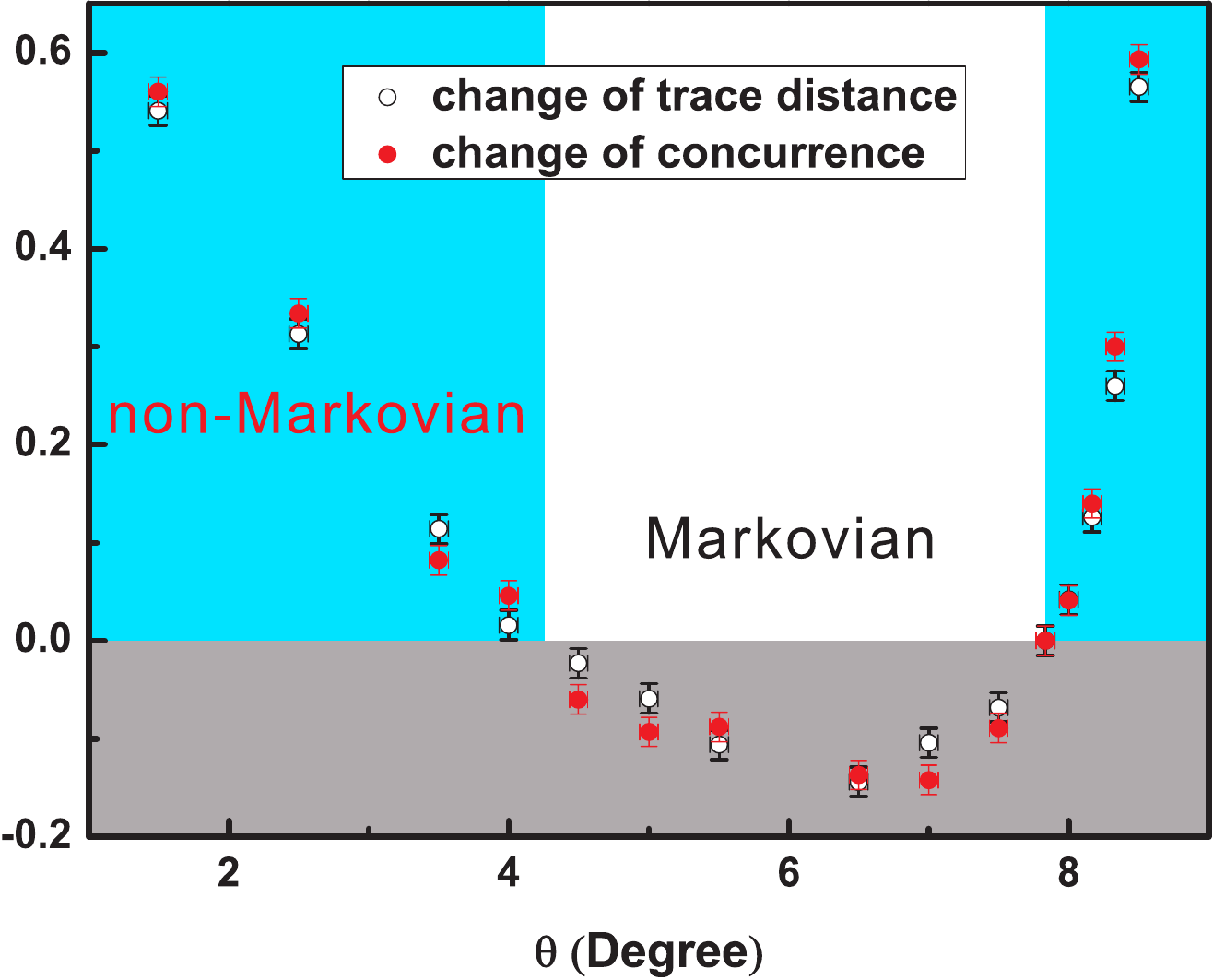}
\caption{\label{Fig:4} The change of the trace distance and of the concurrence as
functions of the tilting angle $\theta$. The transition
from the non-Markovian to the Markovian regime occurs at
$\theta\simeq 4.1^{\circ}$, and from the Markovian to the
non-Markovian regime at $\theta\simeq 8.0^{\circ}$. The
positive values in the blue regions give directly the
non-Markovianity measure ${\mathcal{N}}(\Phi)$ of the
process. The negative values in the grey area correspond to
${\mathcal{N}}(\Phi)=0$, i.e., to Markovian dynamics. 
The error bars are due to the uncertainty of the tilting angle and the counting statistics.
}
\end{figure}

The experiment also enables a direct determination of the measure for non-Markovianity (\ref{MEASURE}). First, we note that the initial pair (\ref{init-pair}) is already optimal in the sense that it yields a maximal increase of the trace distance. The theoretical explanation for this fact is presented in the Supplementary Information.
Second, in our experiment the increase of the trace distance is restricted to a single time interval (see Fig.~\ref{Fig:3}). The non-Markovianity measure (\ref{MEASURE}) is thus obtained by determining the difference of the trace distance between the first local minimum and the subsequent maximum. Our experimental results are shown in Fig.~\ref{Fig:4}. Increasing the tilting angle of the cavity decreases the relative amplitude $A$ between the peaks in the frequency spectrum and thereby reduces the non-Markovianity of the process until a transition to Markovian dynamics occurs. Further increasing the tilting angle amplifies the relative amplitude again and brings the dynamics back to the non-Markovian regime.

In Ref.~\cite{RHP} an alternative measure for non-Markovianity has been proposed which is based on the idea that a Markovian dynamics leads to a monotonic decrease of the entanglement between the open system and an isomorphic ancilla system, while a non-Markovian dynamics induces a temporary increase of the entanglement. One can show (see Supplementary Information) that for the present experiment this measure coincides with (\ref{MEASURE}) if one uses the concurrence \cite{Wootters,Rungta} as an entanglement measure. This fact leads to an alternative and independent method for the measurement of the non-Markovianity by means of our experimental setup. Fixing HWP1 to 22.5 degree, we generate a maximally entangled two-photon state. Photon 1 then passes the quartz plate and the composite final state is analyzed through two-photon state tomography. Experimental results are shown in Fig.~\ref{Fig:3}(b) and Fig.~\ref{Fig:4}, clearly demonstrating the equivalence of both measures for non-Markovianity.

Our experiment clearly reveals the measurability of recently proposed theoretical measures for quantum non-Markovianity which yield important information about the type of quantum noise and about environmental properties, even when the environment is a complex system involving an infinite number of degrees of freedom and is not directly accessible through measurements. Moreover, we have introduced a method for the control of the information flow between the open system and its environment, which opens the possibility of efficiently exploiting memory effects in future quantum technologies \cite{Perdomo}.

{\bf  Acknowledgments} 
This work was supported by the National Fundamental Research Program, National Natural Science Foundation of China  (Grant Nos.~10874162 and 10734060), the Magnus Ehrnrooth Foundation, and the Graduate School of Modern Optics and Photonics.

{\bf Author Contributions}
B.-H.L., L.L., Y.-F.H., C.-F.L., and G.-C.G. planned, designed and implemented the experiments.
E.-M.L., H.-P.B., and J.P. carried out the theoretical analysis and developed the interpretation.
B.-H.L., C.-F.L., E.-M.L, H.-P.B., and J.P. wrote the paper and all authors discussed the contents.

{\bf Author Information}
The authors declare that they have no competing financial interests.
Reprints and permissions information is available online at http://npg.nature.com/reprintsandpermissions. 
Correspondence and requests for materials should be addressed to C.-F.L. or J.P.

\section*{Supplementary Information}

\subsection*{Finding the maximizing pair of initial states}
For an arbitrary initial pair $\rho_{1,2}(0)$ the trace distance evolution is given by
$D(\rho_1(t),\rho_2(t)) = \sqrt{a^2+|\kappa(t)b|^2}$,
where $a=\rho_1^{11}(0)-\rho_2^{11}(0)$ denotes the difference of the initial populations, and $b=\rho_1^{12}(0)-\rho_2^{12}(0)$ the difference of the initial coherences. It follows that any growth of the trace distance is maximal for $a=0$ and $|b|=1$, in which case we obtain the formula (\ref{OPTIMAL-DISTANCE}). Therefore, the pair of states given by Eq.~(\ref{init-pair}) is an optimal initial pair. More generally, all initial pairs of states are optimal which correspond to pairs of antipodal points on the equator of the Bloch sphere that represents the two-state state system.

\subsection*{Equivalence of the two measures}
The initial state is the pure, maximally entangled system-ancilla state $\rho_{\rm SA}(0) = |\psi_{\rm SA}\rangle\langle\psi_{\rm SA}|$ with
$ |\psi_{\rm SA}\rangle = \frac{1}{\sqrt{2}}\left(
 |HH\rangle + |VV\rangle \right).$
The dynamical map acts locally on the system part leading to the state
\begin{eqnarray*}
 \rho_{\rm SA}(t) &=& (\Phi_t\otimes I)\rho_{\rm SA}(0) \\
 &=& \frac{1}{2} \Big(
 |HH\rangle\langle HH| + |VV\rangle\langle VV| \\
 && + \kappa(t)^*|HH\rangle\langle VV| + \kappa(t) |VV\rangle\langle HH| \Big).
\end{eqnarray*}
This state is diagonal in a basis of maximally entangled states (Bell state basis), the maximal eigenvalue being $p_{\max} = \frac{1}{2}(1+|\kappa(t)|)$. Therefore, the concurrence
of $\rho_{\rm SA}(t)$ is given by $C(\rho_{\rm SA}(t)) = 2 p_{\max} - 1 = |\kappa(t)|$ \cite{Hayashi}. Thus, we have the relation
$D(\rho_1(t),\rho_2(t)) = C(\rho_{\rm SA}(t))$, showing that
both measures yield identical results.

\end{document}